# Line Edge Roughness Effects on the Thermoelectric Properties of Armchair Black Phosphorene Nanoribbons

*Ebrahim Pishevar[1], and Hossein Karamitaheri[1,2*]*

[1]Institute of Nanoscience and Nanotechnology, University of Kashan, Kashan 87317-53153, Iran
[2]Department of Electrical and Computer Engineering, University of Kashan, Kashan 87317-53153, Iran
[*]karamitaheri@kashanu.ac.ir

**Abstract:**
This study delves into the thermoelectric properties of armchair black phosphorene nanoribbons while considering the presence of line edge roughness. Employing the tight-binding method in conjunction with non-equilibrium Green's function techniques and Landauer formulas, we explore the impact of various parameters on thermoelectric performance. Our findings reveal that the electrical conductivity and, consequently, the power factor exhibit an increasing trend with expanding ribbon length and width. This behavior can be attributed to heightened collision rates, particularly in narrow ribbons, induced by line edge roughness as length increases. Remarkably, the Seebeck coefficient at the Fermi energy corresponding to the maximum power factor remains nearly constant across different widths, lengths, temperatures and transport regimes. Furthermore, the thermoelectric figure of merit demonstrates a positive correlation with both ribbon length and width. In narrow widths and lengths around 1000 nm, the power factor and figure of merit exhibit an upward trend with ribbon width. However, with further increases in ribbon width, the influence of line edge roughness on thermal conductivity diminishes. Consequently, the figure of merit decreases due to the rise in thermal conductivity. Notably, the thermoelectric figure of merit is higher for short and narrow ribbons and long and wider ribbons.

## 1. Introduction

Thermoelectric materials have garnered significant attention. These materials, in particular, have the capability to directly convert heat into electricity, making them a promising and hopeful choice for applications as clean energy converters.[1-4] Thermoelectric devices made from thermoelectric materials exhibit simple, cost-effective, silent, and durable performance since they lack any moving parts. These devices are reliable and find applications in various industries and specific uses, such as recovering wasted heat in power plants, reducing fuel consumption in vehicles, powering electrical systems in aerospace, wearable thermoelectric devices for utilizing body heat to generate low-power electricity, solar-thermoelectric generators, sensor construction in medical applications, and energy-efficient building facades.[2-9] So far, commercial thermoelectric materials have primarily been based on alloys such as PbTe, Bi2Te3, and SiGe. However, the use of these materials is limited due to their scarcity, toxicity, and high cost.[1, 5, 10-12]

The energy conversion efficiency of thermoelectric materials is primarily assessed by a dimensionless figure of merit (ZT). ZT is inversely proportional to thermal conductivity and directly proportional to the Seebeck coefficient and electrical conductivity. Achieving a minimum ZT value of 4 is essential for competitiveness with other thermal-to-electric conversion technologies. In this context, thermoelectric parameters demonstrate interdependence, presenting a significant challenge in improving each parameter without affecting others.[2, 5, 12-15]

Strategies for improving the performance of thermoelectric materials include various approaches such as doping and alloying approach, heavy element addition, superlattice approach, band engineering approaches, Phonon-Glass Electron-Crystal (PGEC), pressure-induced approach, and nanostructuring approach, among others.[5, 10, 12, 13, 16, 17] Among the most effective methods are nanostructuring approach, including the use of two-dimensional nanostructures, nanotubes, nanowires, and quantum dots etc.[13, 18, 19]

The arrival of materials at the nanoscale leads to changes in their electronic and physical properties due to the emergence of quantum confinement effects. nanostructuring confines the carrier's freedom of movement in specific directions, resulting in significant alterations in electronic transport behavior. In one or two-dimensional nanostructures, phonons are effectively and selectively scattered compared to electrons. Increasing the surface area in this method enhances phonon scattering, significantly reducing lattice thermal conductivity.[5, 10, 16, 20] This can be achieved by creating nanostructures with dimensions smaller than the phonon mean free path but larger than the electron mean free path. Using nanostructures allows the enhancement of the Seebeck coefficient while reducing thermal conductivity, potentially maintaining or even increasing electrical conductivity. These results collectively indicate that nanostructuring serves as a highly effective approach to reduce mutual dependencies between thermoelectric parameters and consequently enhance the thermoelectric figure of merit. This approach emerges as one of the most promising methods for improving the thermoelectric properties of materials.[10, 11, 16]

Two-dimensional nanoribbons possess a plethora of intriguing characteristics including flexibility, a substantial surface-to-volume ratio, and pronounced quantum confinement and edge effects. Consequently, the unique structural attributes of nanoribbons offer precise control over their band structure, the emergence of novel features, and the creation of specialized geometric configurations, all of which hold significant potential for diverse applications.[21-23]

Black phosphorus, a highly stable allotrope of phosphorus first synthesized by Brichman in 1914, shares similarities with graphene in terms of layering and is separable through methods like mechanical exfoliation. Each individual layer, known as phosphorene, is held together by weak van der Waals forces.[24-27] Phosphorene exhibits an anisotropic structure, characterized by ductility along one of its in-plane crystal directions and stiffness along the other, resulting in distinctive thermoelectric, optical, chemical, and unconventional transport properties.[23, 28-30] Among its key attributes are an intrinsic, non-zero band gap and a relatively large mean free path, rendering phosphorene a formidable competitor among two-dimensional materials, such as graphene and silicene. The inherent anisotropy of phosphorene extends to its thermal conductivity, as ab initio calculations have demonstrated.[29-34] Furthermore, this anisotropy in electrical and thermal transport positions phosphorene favorably for thermoelectric applications compared to its counterparts like graphene.[9, 35]

Oxidation of phosphorene, expected to occur spontaneously, can lead to an improvement in the thermoelectric performance compared to its pristine state.[36] The twisted layers' edges in black phosphorene induce strong anharmonic phonon scattering and a significant reduction in lattice thermal conductivity.[37] In a study by Sodagar et al., the effect of strain on the energy gap and edge state scattering, resulting in a twofold increase in the thermoelectric figure of merit of bilayer phosphorene nanoribbons, was investigated.[38]

Zhu et al. have found that applying an electric field to zigzag phosphorene nanoribbons induces even and odd effects on the thermoelectric properties, leading to different behaviors of edge states for even-odd cases.[39] Rezayi et al. have suggested that by creating a periodic structure of vacancies, new states in the energy gap of armchair phosphorene nanoribbons can be introduced, which is effective in optimizing their thermoelectric performance.[40] In addition, Researchers' investigations indicate that the lattice thermal conductivity in zigzag phosphorene nanoribbons can be tunable through chemical doping with materials such as carbon and germanium.[41]

The electrical conductivity and Seebeck coefficient measured in phosphorene nanoribbons demonstrate a clear temperature dependence in both the zigzag and armchair directions.[30] In contrast, the thermal conductivity of phosphorene exhibits a pronounced anisotropy, with the value in the zigzag direction being three times greater than that in the armchair direction.[28] While anisotropy arises from the geometric structure of phosphorene in the case of electrical and thermal conductance, its Seebeck coefficient remains isotropic. In the context of thermoelectric performance, the thermoelectric figure of merit, ZT, exhibits an increase with temperature and showcases significantly larger values in the armchair direction as compared to the zigzag direction.[42]

Despite significant advancements in material fabrication technology, achieving the precise fabrication of narrow nanoribbons from two-dimensional materials with atomic-level precision at their edges remains an unattained goal.[43] A promising avenue for exploring new and intriguing aspects of one and two-dimensional materials involves the investigation of surface and line edge roughness.[44, 45] In essence, edge roughness parameters can be harnessed to tailor specific electronic, thermal, and thermoelectric properties within these materials.

In a study focusing on the thermal conductivity of graphene nanoribbons in the presence of edge roughness, it was demonstrated that line edge roughness stands as one of the most influential factors affecting thermal conductivity.[32] Similarly, the electrical conductivity of graphene nanoribbons is also influenced by line edge roughness.[31, 46] In the case of narrow graphene nanoribbons, wherein the band gap is intimately linked to the ribbon's width, line edge roughness emerges as the dominant collision mechanism. Within these nanoribbons, roughness exerts a more pronounced impact on electrical conductivity compared to thermal conductivity. As a result, high thermoelectric efficiency in armchair graphene nanoribbons is not anticipated.

In our previous work, we elucidated a crucial distinction between graphene nanoribbons and black phosphorene nanoribbons concerning line edge roughness.[47] While in graphene nanoribbons, line edge roughness constrains the effective mean free path of charge carriers to a few tens of nanometers, in black phosphorene nanoribbons, mean free paths can extend up to 2 μm in nanoribbons with a width of approximately 5 nm.[32, 33, 47-49] Notably, in the case of armchair phosphorene nanoribbons with widths exceeding 3 nm, the impact of line edge roughness on electron scattering becomes nearly negligible when compared to electron-phonon scattering effects. Consequently, in armchair phosphorene nanoribbons, line edge roughness exerts a minimal influence on electrical conductivity and may primarily reduce thermal conductivity.[32, 34, 44, 46, 47, 49] This distinctive behavior suggests the potential for relatively high thermoelectric efficiency in armchair phosphorene nanoribbons.[47] Having conducted thorough searches in this domain, no prior studies or research endeavors have been dedicated to examining the influence of edge roughness on the thermoelectric characteristics of phosphorene nanoribbons. Therefore, this investigation marks the inaugural exploration into this particular aspect. It is worth noting that the impact of edge roughness on the thermoelectric properties of other materials, such as nanowires and graphene nanoribbons, has been previously scrutinized.[33, 36, 37, 39-41, 50-56] Additionally, given that alterations in the width of armchair phosphorene nanoribbons do not significantly impact their energy gap, we show in this paper that these materials could emerge as promising candidates for utilization in thermoelectric applications.[47]

In this study, we have conducted a comprehensive investigation into the impact of line edge roughness on the electronic and thermoelectric properties of armchair black phosphorene nanoribbons. The subsequent sections of this article are organized as follows: In Section 2, we detail the computational methods employed and provide an overview of the thermoelectric parameters and their relationships. Section 3 is dedicated to presenting the simulation results and engaging in discussions. Finally, our findings and conclusions are summarized in Section 4.

## 2. Computational Methods

A range of methods and approximations are employed to calculate the band structure of materials. Notable among these are density functional theory (DFT), linear combination of atomic orbitals, GW approximation, and the tight-binding method. DFT and other pseudo-potential-based methods are renowned for their ability to yield highly accurate results compared to alternative computational approaches. Additionally, the tight-binding method stands out due to its capacity for high-speed calculations, making it a suitable choice for structures comprising thousands of atoms, all while offering computational efficiency surpassing that of alternative methods.

The unit cell of phosphorene comprises four phosphorus atoms, denoted as $A.B.A'$. and $B'$ with corresponding positions defined as follows: $\tau_A$=(uc, 0, vb), $\tau_B$=(($\frac{1}{2}$-u)c, $\frac{a}{2}$, vb), $\tau_{A'} = -\tau_A, \tau_{B'} = -\tau_B$. The crystal parameters for phosphorene are characterized by $a = 3,314$ Å. $b = 10,478$ Å. $c = 4,376$ Å. $u = 0,08056$ $and$ $v = 0,10168$. The tight-binding parameters for black phosphorene, considering interactions up to the fifth nearest neighbors, are detailed in **Table 1**.[47] Utilizing these specified values, it is possible to calculate the band structure of black phosphorene, yielding results that are in a good agreement with those obtained via the GW approximation.[47]

**Table 1**: Black phosphorene tight-binding parameters considering up to the fifth nearest neighbors.[47]

| No | Distance [Å] | Tight-binding parameters [eV] |
|---|---|---|
| 1 | 2.22 | -1.22 |
| 2 | 2.24 | 3.665 |
| 3 | 3.34 | -0.205 |
| 4 | 3.47 | -0.105 |
| 5 | 4.23 | -0.055 |

From the band structure and under ideal ballistic conditions, the transmission function can be derived by counting the number of bands at a given energy level.[36] However, in real fabricated structures, transmission is typically not ballistic, and some scattering occurs during the electron transfer through the channel. These scattering events can be attributed to various mechanisms, including atomic defects and line edge roughness, which may arise during the fabrication process. In the context of this study, we exclusively consider line edge roughness as a dominant scattering mechanism in narrow nanoribbons, potentially transitioning carriers from a ballistic to a non-ballistic regime. Across all structures examined in this study, we assume an rms roughness amplitude of 0.2 nm and a roughness correlation length of 2 nm.[32] The line edge roughness is simulated using a random process with the autocorrelation function defined as:

$$R(x) = \Delta w^2 \exp\left(-\frac{|x|}{\Delta L}\right) \tag{1}$$

Here, $x$-axis is along the length direction of the structure, $\Delta L$ is the correlation length (a measure of smoothness), and $\Delta w$ is the square root of the mean square roughness amplitude.

To generate the line edge roughness in real space, the Fourier transform of the autocorrelation function is computed, resulting in the power spectrum. By introducing a random phase to the power spectrum and subsequently performing the inverse Fourier transform, the roughness profile in real space is derived. 50 samples are generated with identical roughness parameters, and their transmission functions are determined. Finally, a statistical average of the total transmission functions is calculated to obtain the ultimate transmission function. [46, 57]

In non-ballistic regimes, one can employ the non-equilibrium Green's function technique in order to obtain the transmission function. To facilitate this, a structure composed of two contacts and a channel between them is employed. To investigate the impact of line edge roughness on channel characteristics, we assume that the contacts are free of roughness and in equilibrium. **Figure 1** illustrates the structure of a representative armchair phosphorene nanoribbon in the presence of line edge roughness. Red lines have been used to seperate contacts and channel parts. Green lines have been used to show the ribbon's width before introducing roughness on the channel. The Hamiltonian of the channel, the contacts, and the overlap of the channel with the contacts are all determined using the tight-binding method. Subsequently, the Green's function is obtained, following the approach used in prior studies.[18, 46, 48, 57]

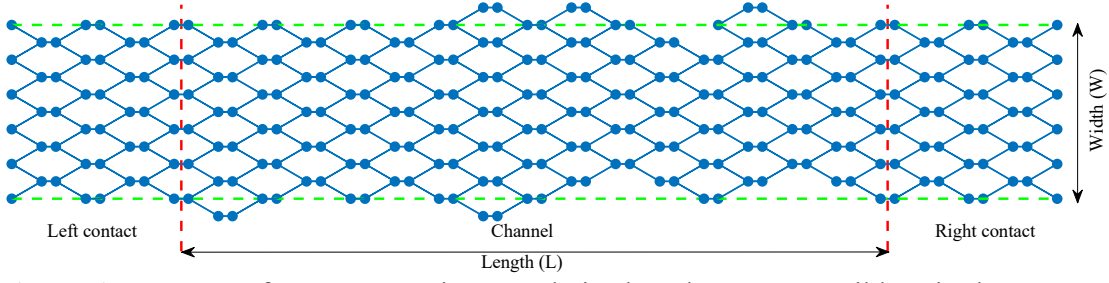

**Figure 1** Structure of a representative armchair phosphorene nanoribbon in the presence of line edge roughness. Red lines have been used to seperate contacts and channel parts. Green lines have been used to show the ribbon's width before introducing roughness on the channel.

## 2.1 The Figure of Merit and Landauer Relations

The figure of merit (ZT), which directly represents the thermoelectric efficiency of the material, is a unitless positive value calculated as:

$$ZT = \frac{GS^2 T}{K_e + K_p} = \frac{\sigma S^2 T}{\kappa_e + \kappa_p} \qquad (2)$$

In this expression, $G$ is the electrical conductance, $S$ the Seebeck coefficient, $T$ the temperature and $K_e$ and $K_p$ are thermal conductance of electrons and phonons, respectively. $\sigma$ and $\kappa$ are also considered as electrical and thermal conductivity. The term $GS^2$ (or $\sigma S^2$) is called power factor. Each of the constituent parameters in this relation is related to length, width, and temperature.

The Landauer relation is employed to obtain parameters such as electrical conductance, Seebeck coefficient, and thermal conductance. For small voltage differences, one can write:

$$I = \frac{2q^2 V}{h} \int_{-\infty}^{+\infty} T_{el}(E)\left(-\frac{\partial f}{\partial E}\right) dE \qquad (3)$$

where $I$ is the electrical current, $T_{el}(E)$ is the electron transmission function, $f$ is the Fermi distribution function, $E$ is the energy, $h$ is Planck's constant and $q$ is the elementary charge. As a result, electrical conductance can be defined as:

$$G = \frac{2q^2}{h} \int_{-\infty}^{+\infty} T_{el}(E)\left(-\frac{\partial f}{\partial E}\right) dE \qquad (4)$$

The thermal broadening function, denoted as $F_T$, is defined as:

$$F_T = -\frac{df}{dE} = \frac{1}{4k_B T} \operatorname{sech}^2 \frac{E}{2k_B T} \qquad (5)$$

Here, $k_B$ is Boltzmann's constant.

The Seebeck coefficient ($S$) can be calculated as:

$$S = \left(\frac{-K_B}{q}\right) \frac{\int_{-\infty}^{+\infty} T_{el}(E)\left[\frac{E - E_F}{k_B T}\right]\left(-\frac{\partial f}{\partial E}\right) dE}{\int_{-\infty}^{+\infty} T_{el}(E)\left(-\frac{\partial f}{\partial E}\right) dE} \qquad (6)$$

In which $E_F$ represents the Fermi energy. For the electron thermal conductance ($K_e$), it can be expressed as:

$$K_e = K_0 - TS^2 G \qquad (7)$$

where:

$$K_0 = \frac{2}{hT} \int_{-\infty}^{+\infty} T_{el}(E)(E - E_F)^2 \left(-\frac{\partial f}{\partial E}\right) dE \qquad (8)$$

## 2.2 Phonon Thermal Conductance and Mean Free Path

In non-ballistic transmission, within a resistive medium, phonons also experience scattering events that change their transmission. The relationship between phonon thermal conductance in the ballistic transport regime ($K_B$) and the non-ballistic transfer regime ($K_p$), effective mean free path ($\lambda_{eff}$), and the channel length ($L$) is given by:

$$K_p = K_B \frac{\lambda_{eff}}{\lambda_{eff}+L} \tag{9}$$

The thermal conductivity of phonons ($\kappa_p$) can be calculated using the relation:

$$\kappa_p = K_p \frac{L}{A} = \frac{K_B \lambda_{eff}}{L+\lambda_{eff}} \frac{L}{A} \tag{10}$$

where: $A$ is the cross-sectional area of the nanoribbon, with $A = H \times W$, where $W$ is the width of the ribbon and $H$ is its effective thickness. For phonons, $\lambda_{eff}$ is influenced by different collision centers and can be expressed as:

$$\frac{1}{\lambda_{eff}} = \frac{1}{\lambda_{ph-ph}} + \frac{1}{\lambda_R} \tag{11}$$

where $\lambda_{ph-ph}$ and $\lambda_R$ are phonon-phonon dominated and line edge roughness dominated mean free path for phonons, respectively.

## 3. Simulation Results and Discussions

In this section, we delve into the simulation results pertaining to the electrical and thermoelectric characteristics of armchair black phosphorene nanoribbons. We explore the influence of various engineering parameters, including ribbon width, length, and temperature, both in the ballistic regime and in the presence of line edge roughness.

One distinctive feature of black phosphorene nanoribbons is their less sensitivity of the energy band gap to changes in ribbon width. Unlike graphene nanoribbons, where the band gap exhibits significant variations with width, black phosphorene nanoribbons display relatively no alterations in this regard.[47] Consequently, the transmission function and, thus the electrical properties are expected not to be affected much even in the presence of line edge roughness. The following subsections will provide a comprehensive analysis of our simulation results, offering valuable insights into the electrical and thermoelectric behaviors of armchair black phosphorene nanoribbons under different conditions.

### 3.1 Transmission Function

In this subsection, we present the electron transmission function of armchair phosphorene nanoribbons (**Figure 2**). We examine two distinct ribbon widths, namely 2 nm (Figure 2-a) and 5 nm (Figure 2-b), under the conditions of the ballistic regime. Additionally, we investigate the impact of line edge roughness on the transmission function at various ribbon lengths.

As illustrated in Figure 2, for the narrower ribbon with a width of 2 nm, the transmission function exhibits a pronounced feature, a region of zero transmission within the energy range spanning approximately from -0.85 eV to +0.85 eV. This energy range corresponds to the band gap of about 1.7 eV, and it signifies the absence of electron transport within this specific energy range. Beyond the band gap, which occurs at energies greater than approximately +0.85 eV, we encounter the conduction bands. Conversely, at energies less than -0.85 eV, we observe the valence bands, where the transmission function has non-zero values. Furthermore, as we traverse both the positive and negative energy directions, secondary and higher sub-bands emerge, leading to an increase in the value of the transmission function. It is worth mentioning that the band gap of the wider ribbon with a with of 5 nm (Figure 2-b) has a lower band gap of about 1.55eV. If fact, lower ribbons have higher band gap due to the quantum confinement effects.

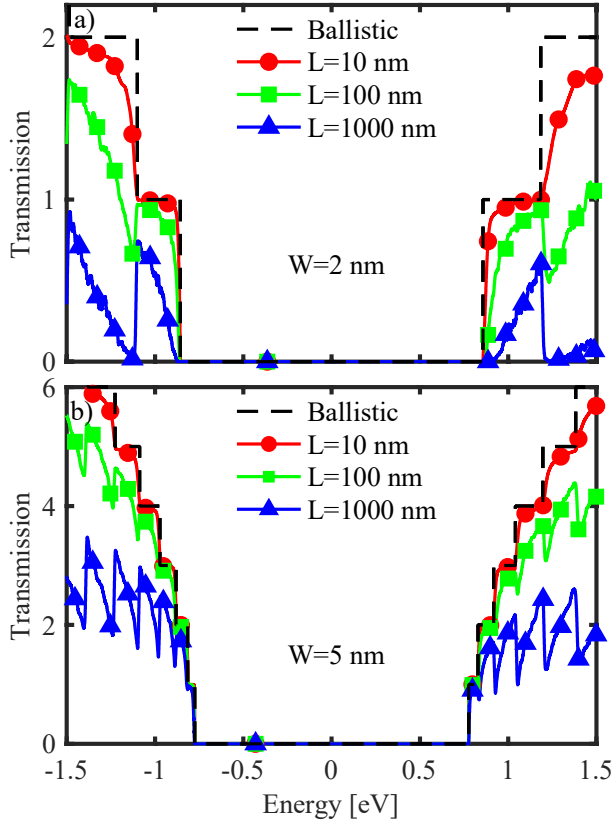

**Figure 2** Transmission function of armchair phosphorene nanoribbons for widths a) 2 nm and b) 5 nm in the ballistic regime and at different lengths in the presence of edge roughness.

In the presence of line edge roughness and under non-ballistic conditions, the transmission function undergoes a reduction. This decline can be attributed to the collisions experienced by carriers as they interact with the ribbon's edges. Particularly, in proximity to the edge of each sub-band, a sharp drop in the non-ballistic transmission function is observed. This phenomenon arises due to the elevated density of states near the edges of sub-bands, which, in turn, results in increased scattering events for electrons. Consequently, the overall transmission function experiences a decrease. Notably, this effect becomes more pronounced with an increase in the ribbon's length, as longer ribbons provide additional opportunities for collision events.

Furthermore, a noteworthy observation is that the energy separation between the edges of the first and second sub-bands is greater in the narrower ribbon compared to the wider ribbon. In other words, in a wider ribbon, sub-bands manifest with smaller energy differences. Assuming that the Fermi energy resides near the band edge, this leads to a notable overlap between the thermal broadening function and the transmission function. Consequently, the electrical conductance is augmented, potentially influencing the thermoelectric parameters.

In the context of ribbon width, the wider ribbon (as depicted in Figure 2-b) shows a smaller band gap compared to the narrower one. This result aligns with the anticipated inverse relationship between the band gap and nanoribbon width. In the wider ribbon, the transmission function values are higher, signifying a greater number of bands at a given energy level in comparison to the narrower ribbon. Additionally, the wider ribbon displays a reduced susceptibility to the impact of roughness on the transmission function. This reduced impact is attributed to the fact that edge roughness affects a smaller fraction of the ribbon's total width in wider ribbons.

## 3.2 Ballistic regime

In the ballistic transport regime, as depicted in **Figure 3**, we examine the electrical conductance (Figure 3-a), Seebeck coefficient (Figure 3-b), and power factor (Fig 3-c) for an armchair phosphorene nanoribbon with a width of 2 nm.

The electrical conductance is relatively low when the Fermi energy is positioned in the middle of the band gap. However, as the Fermi energy approaches the edge of the conduction band, the electrical conductance experiences an increase. This behavior can be attributed to the growing overlap between the thermal broadening function and the transmission function as the Fermi energy approaches the edge of the conduction band, leading to an increase in electrical conductance. Additionally, the electrical conductance rises with increasing temperature. This increase is a consequence of the thermal broadening function expanding as temperature rises, resulting in a greater overlap with higher values of the transmission function. Essentially, the increase in temperature enables the involvement of additional sub-bands in electrical conduction.

The behavior of the Seebeck coefficient in armchair phosphorene nanoribbons is particularly interesting. At the Fermi energy near the middle of the band gap, the Seebeck coefficient effectively approaches zero due to the neutralization of the contributions of electrons and holes. However, with a slight change in the Fermi energy and the emergence of an asymmetry between electron and hole conduction, the Seebeck coefficient reaches its maximum value. It is worth mentioning that although transmission function is zero in the energy range of band gap, since there is an integral expression covering all energies from negative infinity to positive infinity, the denominator integral of the Seebeck coefficient (equation 6) will not be zero due to the contribution of all values of $T_{el}(E)$ across all energies. Therefore, the Seebeck coefficient will not diverge. As the Fermi energy continues to increase, resulting in an increase in electrical conductance (which has an inverse relationship with the Seebeck coefficient) and therefore the Seebeck coefficient decreases once again.

At lower temperatures, the maximum value of the Seebeck coefficient is approximately 9 mV/K, but as the temperature increases to 600 K, this value decreases to about 1 mV/K. This decline is primarily due to the increase in electrical conductance with temperature. One noteworthy observation is that the maximum values of the Seebeck coefficient do not align with the Fermi energy corresponding to the maximum values of the power factor. The Fermi energy for maximum power factor values is located a few $k_B T$ smaller than the edge of the conduction band. This seeming discrepancy can be attributed to the very small values of electrical conductance at Fermi energies corresponding to the maximum values of the Seebeck coefficient. The symbols in Figure 2 represent values corresponding to the maximum power factor at various temperatures, providing a visual reference for these key points.

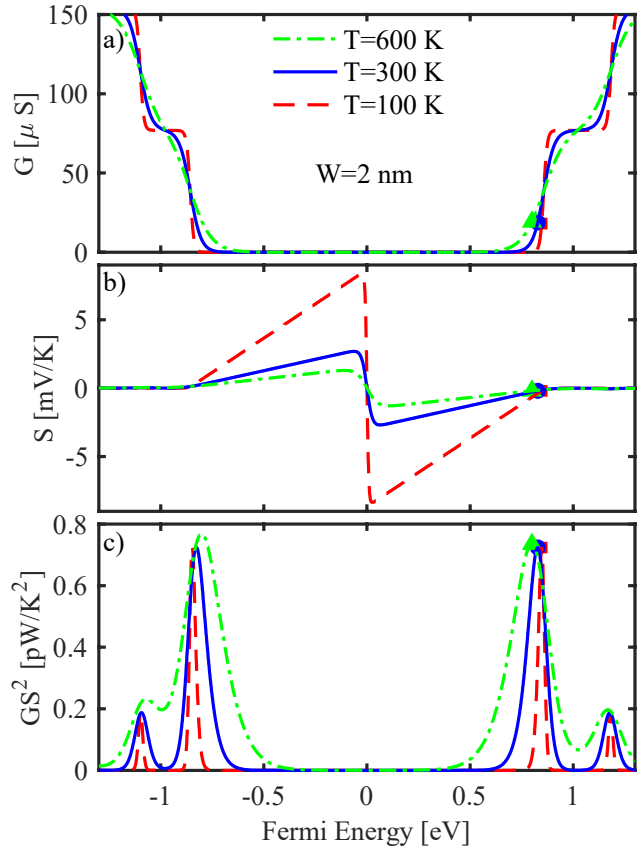

**Figure 3** a) Electrical conductance, b) Seebeck coefficient and c) power factor for an armchair phosphorene nanoribbon with a width of 2 nm in the ballistic transport regime.

In **Figure 4**, the electrical conductance (Figure 4-a), Seebeck coefficient (Figure 4-b), and power factor (Figure 4-c) are plotted as functions of temperature for armchair phosphorene nanoribbons with widths of 2 nm and 5 nm, considering Fermi energies that correspond to the maximum power factor values. For the narrow ribbon (W=2 nm), it is observed that increasing the temperature has a negligible effect on the electrical conductance. This implies that the electrical conductance at Fermi energies corresponding to the maximum power factor values remains relatively constant at different temperatures. However, in the wider ribbon (W=5 nm), the electrical conductance is significantly higher, especially at high temperatures. The increase in electrical conductance with temperature in wider ribbons can be attributed to the larger values of the transmission function compared to narrow ribbons.

In contrast to electrical conductance, the Seebeck coefficient values appear to be independent of ribbon width and exhibit minimal changes with increasing temperature. Specifically, the values of the Seebeck coefficient at different temperatures, corresponding to Fermi energies that maximize the power factor, remain at approximately $0{,}2 \text{ mV}K^{-1}$. This consistency is noteworthy, considering that the Seebeck coefficient near zero Fermi energy exhibits significantly higher values.

It's important to note that the presented values in Figure 4 correspond to the Fermi energies where the power factor reaches its maximum. As a result, while the maximum Seebeck coefficient is highly dependent on temperature and ribbon width, its impact on the peak of the power factor is limited. Consequently, the power factor largely follows the behavior of electrical conductance, benefiting from the higher electrical conductance in wider ribbons to achieve higher values.

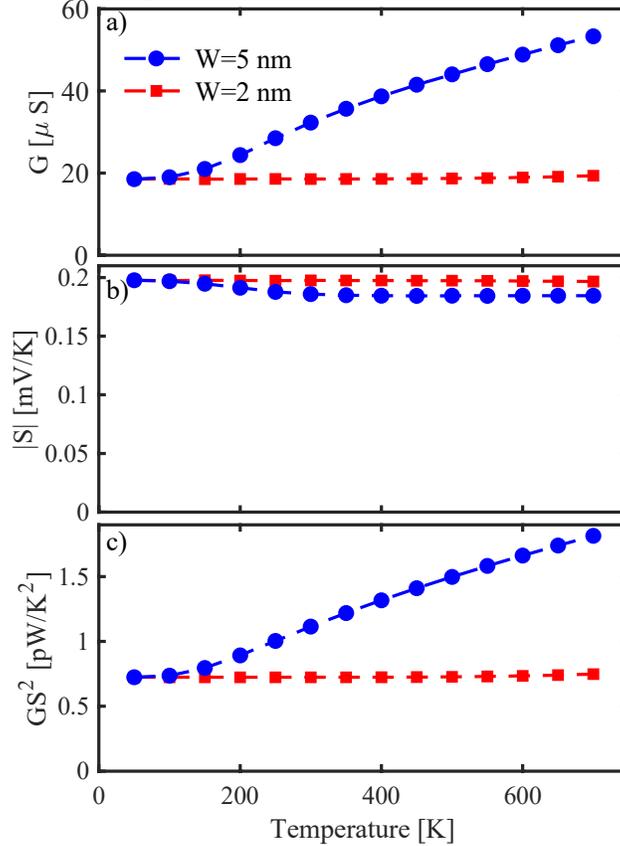

**Figure 4** a) Ballistic electrical conductance, b) Seebeck coefficient and c) power factor in terms of different temperatures at Fermi energy in which the power factor has its maximum value. The results are given for the widths of 2 nm and 5 nm for armchair phosphorene nanoribbon.

In **Figure 5**, the electrical conductance (Figure 5-a), Seebeck coefficient (Figure 5-b), and power factor (Figure 5-c) are depicted. In this figure, for each ribbon width, Fermi energy value is chosen to be corresponding to the maximum power factor at room temperature. This choice is practical for real devices where the Fermi energy typically remains constant and does not change with temperature within the operating temperature range. Consequently, the results in this figure differ from those in Figure 4, particularly at low temperatures, which are less relevant. Furthermore, thermoelectric generators are commonly used at temperatures higher than room temperature, making variations at low temperatures less important.

For both 2 nm and 5 nm ribbon widths, the electrical conductance values increase with temperature. This behavior is attributed to the broadening of the thermal broadening function at higher temperatures, leading to more significant overlap with the transmission function and, consequently, higher electrical conductance values. The electrical conductance is consistently higher for the wider ribbon, and the increase in electrical conductance with temperature is more pronounced in the wider ribbon. This effect can be explained by the smaller band gap in the wider ribbon.

As shown in Figure 5-b, the Seebeck coefficient reaches higher values at lower temperatures due to the decrease in electrical conductance with increasing temperature.

This means that the Seebeck coefficient sharply decreases up to temperatures about room temperature, eventually stabilizing at an approximate value of $0{,}2\ mVK^{-1}$. A comparison between Figure 4-b and Figure 5-b reveals that, at temperatures above room temperature, the Seebeck coefficient values are nearly the same.

Due to the significantly higher electrical conductance values in the wider ribbon compared to the narrower ribbon, the power factor values for the wider ribbon are greater, and they increase with temperature in the case of the wider ribbon. Interestingly, at low temperatures, despite the decreasing Seebeck coefficient values with increasing temperature, both ribbons maintain an upward trend in the power factor in the selected Fermi energies.

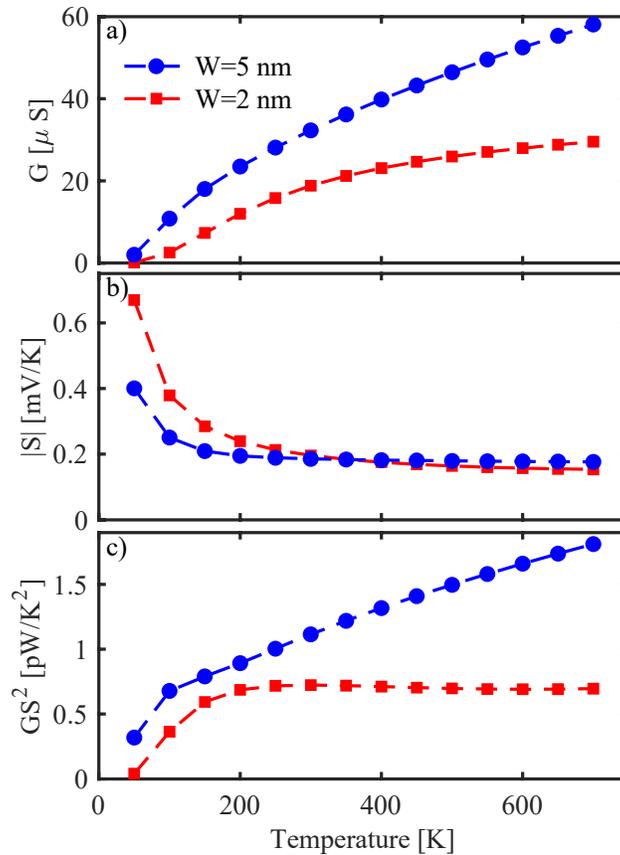

**Figure 5** a) Ballistic electrical conductance, b) Seebeck coefficient and c) power factor, in terms of temperature and for armchair phosphorene nanoribbons of widths 2 nm and 5 nm. The values are presented at the Fermi energies corresponding to the maximum power factor at room temperature.

**Figure 6** illustrates the values of electrical conductivity (Figure 6-a), Seebeck coefficient (Figure 6-b), and power factor (Figure 6-c) as functions of different channel lengths in the presence of line edge roughness for two ribbon widths of 2 nm and 5 nm. Similar to Figure 5. a Fermi energy value corresponding to the maximum power factor at room temperature has been selected for both ribbon widths, and the thermoelectric parameters at that Fermi energy are plotted as functions of channel lengths.

With an increase in the length of the structures, there is a substantial increase in electrical conductivity for both nanoribbons, and this increase is nearly exponential. The steepness of the curves is much higher for lengths longer than 100 nm compared to lengths shorter than 100 nm. Furthermore, at lengths longer than approximately 100 nm, the electrical conductivity of the wider ribbon surpasses that of the narrower ribbon.

The values of the Seebeck coefficient remain relatively stable with variations in channel length, decreasing slightly only for longer lengths in the case of narrow ribbons as the length of the nanoribbon increases. Both structures, with widths of 2 nm and 5 nm, maintain an almost constant value of approximately $0,2\ mVK^{-1}$ for the Seebeck coefficient, consistent with the observations in the ballistic transport regime for ribbons of different widths and temperatures. Interestingly, changes in width, length, and temperature have not had a significant impact on the Seebeck coefficient values corresponding to the Fermi energy that maximizes the power factor. The power factor values follow the increasing trend of electrical conductivity for both ribbon widths of 2 nm and 5 nm, primarily due to the nearly constant values of the Seebeck coefficient. Wider ribbons exhibit higher electrical conductivity, resulting in a greater power factor. Therefore, it appears that using wider and longer ribbons in the design of thermoelectric devices can lead to achieving a higher power factor and a better thermoelectric figure of merit.

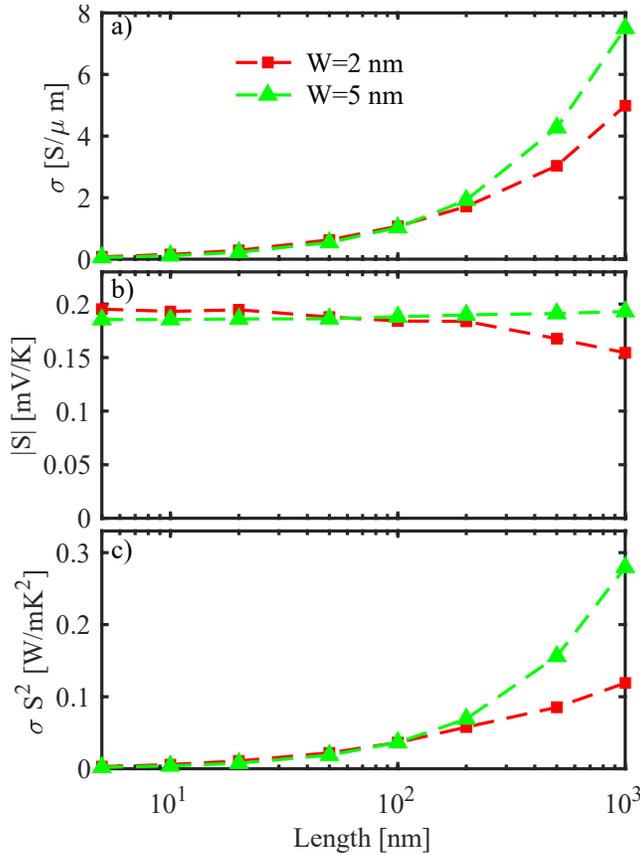

**Figure 6** a) Electrical conductivity, b) Seebeck coefficient and c) power factor as a function of cannel length in the presence of line edge roughness, for two ribbons with widths of 2 nm and 5 nm and at room temperature.

The thermal conductivity ($\kappa$) of armchair black phosphorene nanoribbons in the presence of line edge roughness is given by the equation $\kappa = \kappa_e + \kappa_p$, where $\kappa_e$ is calculated using equation (7). To obtain $\kappa_p$, one can use equation (10) along with the data provided in **Table 2**. Ballistic phonon thermal conductance ($K_B$) for armchair phosphorene nanoribbons with different widths, extracted from reference[58], is also listed in Table 2. Additionally, the phonon-phonon collisions dominated phonon thermal conductivity ($\kappa_{p.ph-ph}$) for the phosphorene sheet in the armchair direction, is reported to be approximately 24 WK$^{-1}$m$^{-1}$ in reference.[59]

one can use these values in equation (10) for long lengths ($L \gg \lambda_{ph-ph}$) as follows: $\kappa_{p.ph-ph} = \frac{K_B \lambda_{ph-ph}}{L+\lambda_{ph-ph}} \frac{L}{A} \cong \frac{K_B}{A} \lambda_{ph-ph}$, in order to obtain the mean free path due to phonon-phonon collision ($\lambda_{ph-ph}$) to be approximately 33.33 nm. In these calculations, $\frac{K_B}{A}$ is considered to be 0,72 nWK$^{-1}$nm$^{-2}$ from reference [58] and a thickness ($H$) of 0,55 nm from reference [47] for the phosphorene sheet.

In nanoribbons, phonon transport is typically not ballistic, and some scattering occurs due to various mechanisms, including edge roughness and phonon-phonon collisions. Equation (11) takes into account the effects of both types of mechanisms, yielding the effective mean free path ($\lambda_{eff}$), which combines the mean free path due to phonon-phonon collisions ($\lambda_{ph-ph}$) and the mean free path due to edge roughness ($\lambda_R$). The values for $\lambda_{eff}$ for armchair phosphorene nanoribbons in the presence of edge roughness are provided in Table 2 and were calculated using data presented in reference[18] and using equation (11). It's worth noting that these values are calculated based on the assumption that the effect of edge roughness on phonons in phosphorene nanoribbons is approximately similar to that in graphene nanoribbons. In relatively long lengths, the thermal conductivity of electrons ($\kappa_e$) dominates $\kappa$, and the approximation used for $\lambda_R$ and $\kappa_p$ is good enough without significantly affecting the results; trends.

**Table 2**: Ballistic phonon thermal conductance, K$_B$. for armchair phosphorene nanoribbons with widths of 1 to 5 nm and mean free path, $\lambda_{eff}$. for armchair phosphorene nanoribbons in the presence of line edge roughness.

| $W$ [nm] | $K_B$ [nWK$^{-1}$] Taken from[58] | $\lambda_{eff}$ [nm] Evaluated using equation (11) |
|---|---|---|
| 1 | 0.3 | 1.57 |
| 2 | 0.65 | 3.57 |
| 3 | 1.1 | 6.89 |
| 4 | 1.6 | 8.98 |
| 5 | 2.00 | 10.66 |

**Figure 7** illustrates the thermal conductivity (Figure 7-a) and the thermoelectric figure of merit, ZT (Figure 7-b), for phosphorene nanoribbons with widths ranging from 2 nm to 5 nm across various channel lengths. As observed, the rising trend seen in power factor and electrical conductivity in Figure 6 is similarly mirrored in thermal conductivity across all ribbon widths. A significant point in Figure 7-a is the dominance of electron thermal conductivity over phonon thermal conductivity, particularly in longer ribbon lengths. Consequently, the approximations employed for κ calculations are not expected to introduce substantial errors into the results. Encouragingly, wider phosphorene nanoribbons exhibit an increasing trend in thermoelectric figure of merit with extended structure lengths. This rise can be attributed to the steeper increase in the power factor compared to thermal conductivity. The obtained plots for the thermoelectric figure of merit in this figure exhibit different behaviors depending on the dominance of each parameter involved in the ZT relation for various lengths and widths. To illustrate, first, the plot for a nanoribbon with a width of 2 nm is examined. As observed, for structure lengths (up to approximately 200 nm), the power factor (numerator of the ZT expression) dominates, resulting in an increase in the thermoelectric figure of merit with increasing length. Meanwhile, for longer lengths, the effect of thermal conductivity ($\kappa$) becomes dominant, causing a decrease in the plot with increasing structure length (for lengths greater than approximately 200 nm). Our estimation for the critical size of the thermoelectric figure of merit suggests that, for longer lengths, relatively wider widths will yield the highest values of the figure of merit (specifically at widths of 5, 6 nm). We expect that by using a wider and longer ribbon one can achived a ZT value of higher than even 3.

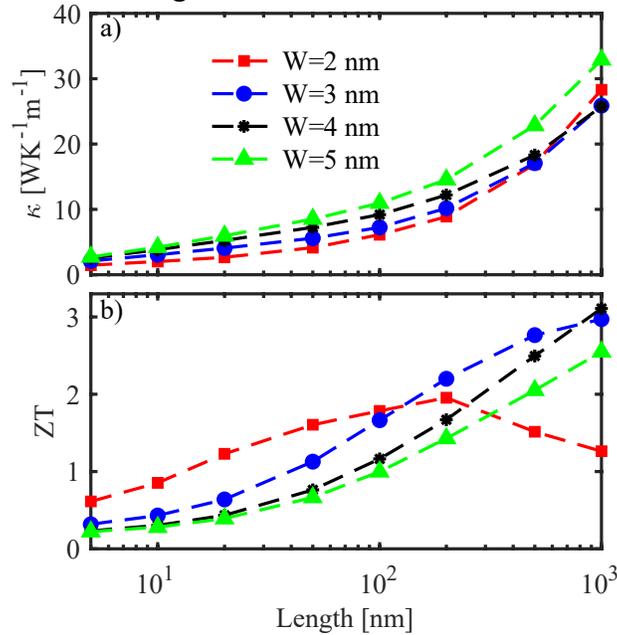

**Figure 7** a) Thermal conductivity $\kappa$ and b) thermoelectric figure of merit, ZT, for nanoribbons with widths of 2 nm up to 5 nm in terms of different channel lengths and at room temperature.

**Figure 8** illustrates the thermal conductivity (Figure 8-a) and the thermoelectric figure of merit, ZT (Figure 8-b), for armchair phosphorene nanoribbons in the presence of edge roughness, considering various lengths and ribbon widths. It is observed that the thermal conductivity values for shorter nanoribbons exhibit a slight increase with the widening of the ribbon. For lengths close to 1000 nm, the thermal conductivity values initially experience a slight decline as the ribbon width increases. However, with further increases in ribbon width, the trend reverses, with thermal conductivity values rising alongside the ribbon width. This behavior can be attributed to the heightened influence of edge roughness on the thermal conductivity of narrower nanoribbons, and as the ribbon width increases, the impact of edge roughness diminishes. Furthermore, in line with the findings presented in Figure 7-a, longer nanoribbons exhibit higher thermal conductivity values.

The values of the thermoelectric figure of merit for short nanoribbons remain below unity and decrease with increasing ribbon width. As the ribbon length extends to around 1000 nm, it becomes possible to achieve figure of merit values exceeding one. When dealing with lengths close to 1000 nm, the trend in the figure of merit changes with an increase in ribbon width. Specifically, for narrower ribbon widths, the figure of merit increases with the widening of the ribbon. This observation suggests that in narrower ribbons, the effect of edge roughness significantly impacts electrical conductivity and subsequently enhances the power factor with increased width. In contrast, for ribbon widths of 3 nm and 4 nm, the influence of edge roughness on electrical conductivity and power factor diminishes, resulting in a reduced slope in the variation of the figure of merit. For widths exceeding 4 nm, the figure of merit decreases as a consequence of the elevated thermal conductivity, attributed to the reduced influence of edge roughness. In summary, narrow ribbons experience a greater impact of edge roughness on the thermal conductivity, making the power factor the dominant factor influencing the figure of merit. In wider ribbons, the reduced effect of edge roughness leads to increased thermal conductivity, making thermal conductivity a more influential factor in the figure of merit. As a result, ribbon widths of approximately 3 nm and 4 nm exhibit almost maximum figure of merit values, particularly for lengths close to 1000 nm. Longer ribbons with widths around 4 nm are expected to yield even higher figure of merit values, especially for lengths exceeding 1000 nm.

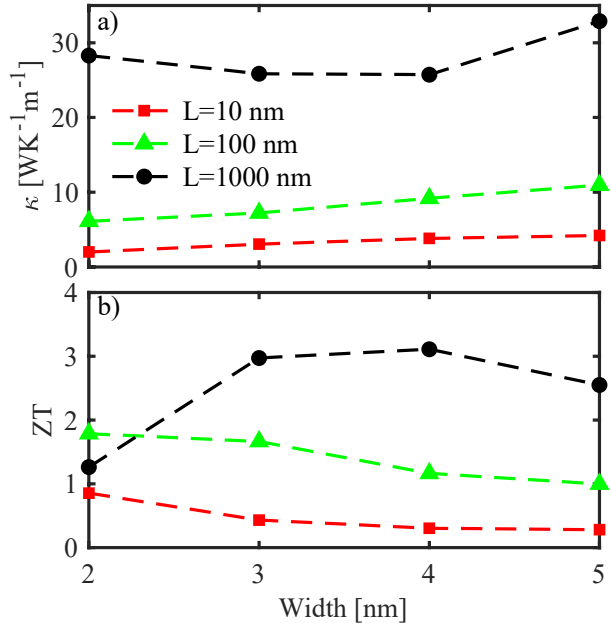

**Figure 8** a) thermal conductivity and b) thermoelectric figure of merit, ZT, as a function of ribbon widths for roughed armchair phosphorene nanoribbons.

## 4. Conclusion

This study delved into simulations of armchair black phosphorene nanoribbons in the ballistic regime, considering the influence of line edge roughness. The research focused on understanding how changes in length, width, and temperature affect the thermoelectric parameters of these nanoribbons.

In the ballistic transport regime, it was observed that wider ribbons (5 nm) exhibit larger transmission function values compared to their narrower counterparts (2 nm). An increase in temperature allowed additional sub-bands to participate in electrical conductance. Interestingly, the study found that the maximum values of the Seebeck coefficient didn't coincide with the Fermi energies associated with the peak power factor; instead, they were slightly shifted, typically a few $k_{BT}$ lower than the conduction band edge. This implies that the values of the Seebeck coefficient vary only slightly with alterations in ribbon length, width, and temperature.

The investigation further revealed that even the effective Seebeck coefficient showed minimal changes in the presence of line edge roughness. In the non-ballistic regime, an increase in ribbon length and width led to enhanced electrical conductivity and, consequently, an increased power factor. This phenomenon is attributed to the more significant impact of collisions in long and narrow ribbons. Promisingly, the figure of merit also exhibited a positive correlation with increases in ribbon length and width.

For nanoribbons with narrow widths (2 nm to 3 nm) and lengths close to 1000 nm, both the power factor and the figure of merit increased with ribbon width. However, in wider ribbons (4 nm and 5 nm), the reduction in the effect of line edge roughness led to increased thermal conductivity with increasing widths, causing a subsequent reduction in the thermoelectric figure of merit.

In conclusion, widths of 3 nm and 4 nm, with lengths close to 1000 nm, demonstrated the highest figure of merit values. It is anticipated that even higher figure of merit values can be achieved for lengths exceeding 1000 nm and widths of around 5 nm. These results substantiate the potential use of roughened black phosphorene nanoribbons, in the presence of edge roughness, in designing efficient and cost-effective thermoelectric devices.

This conclusion underscores the promise of employing black phosphorene nanoribbons, particularly those incorporating edge roughness, in the development of highly efficient and economically viable thermoelectric devices.